\documentclass[
showpacs,nofootinbib
,twocolumn  
]{revtex4-2}

\usepackage{graphicx}
\usepackage{dcolumn}
\usepackage{bm}
\usepackage{epsfig}
\usepackage{amssymb} 
\usepackage{color}
\usepackage{datetime}
\usepackage{slashed} 
\usepackage[mathscr]{euscript}
\usepackage{cancel}
\usepackage[hidelinks]{hyperref} 
\usepackage{microtype}

\usepackage{amsmath}
\newcommand{\bal}{\begin{align}}
\newcommand{\eal}{\end{align}}
\newcommand{\beq}{\begin{eqnarray}}
\newcommand{\eeq}{\end{eqnarray}}
\newcommand{\nneeq}{\nonumber \end{eqnarray}}

\newcommand{\nn}{\nonumber \\} 
\newcommand{\es}{& = &}

\newcommand{\ps}{& + &}
\newcommand{\ms}{& - &}
\newcommand{\ts}{& \times &}
\newcommand{\nt}{\nn \ts}
\newcommand{\np}{\nn \ps}
\newcommand{\nm}{\nn \ms}

\newcommand{\cA}{ {\cal A} }
\newcommand{\cB}{ {\cal B} }

\newcommand{\cC}{ {\cal C} }
\newcommand{\cD}{ {\cal D} }
\newcommand{\cM}{ {\cal M} }
\newcommand{\cH}{ {\cal H} }

\newcommand{\cG}{ {\cal G} }
\newcommand{\cK}{ {\cal K} }
\newcommand{\cT}{ {\cal T} }
\newcommand{\cV}{ {\cal V} }

\newcommand{\cU}{ {\cal U} }

\newcommand{\cY}{ {\cal Y} }

\newcommand{\h}{ {1 \over 2} }

\newcommand{\bmat}{\left[\begin{array}}
\newcommand{\emat}{\end{array}\right]}

\begin{document}

\title{ Cancellation of small-$x$ divergences in the three-gluon-vertex Hamiltonian \\ with canonical gluon mass }
\author{   Juan Jos\'e G\'alvez-Viruet }
\email{juagalve@ucm.es}
\affiliation{ Departamento de F\'isica Te\'orica \\ Universidad Complutense de Madrid }
\author{ María Gómez-Rocha }
\email{mgomezrocha@ugr.es}
\affiliation{  Departamento de F\'isica At\'omica, Molecular y Nuclear \\ and  Instituto Carlos I de Física Teórica y Computacional, Universidad de Granada        }
\date{         \today;\, at \currenttime    }

\begin{abstract}
The front form of Hamiltonian dynamics provides a framework within QCD in which interaction terms are invariant under 7 of 10 Poincar\'e transformations and the vacuum structure is simple. 
However, canonical expressions are divergent and must be regulated before attempting to define an eigenvalue problem.
The renormalization group procedure for effective particles (RGPEP) provides a systematic way of renormalizing Hamiltonians and obtaining counterterms. 
One of its achievements is the description of asymptotic freedom with a running coupling defined as the coefficient of the three-gluon-vertex operators in the renormalized Hamiltonian. 
Yet, the results we obtain need a deeper understanding since the coefficient function shows a finite cutoff dependence, at least in the third-order terms of the perturbative expansion. 
In this work, we present an RGPEP computation of the three-gluon vertex with a different regularization scheme based on massive gluons. 
Our calculation shows that the three-gluon Hamiltonian interaction term has a finite limit as the gluon mass vanishes, but the finite function $h(x)$ that was obtained in previous calculations as a consequence of the finite dependence on the regularization is different. 
This result indicates a need for understanding how to eliminate finite regularization effects from Hamiltonians for effective quarks and gluons in QCD.
Nevertheless, it is remarkable that all terms depending on the gluon mass cancel out in the limit of vanishing gluon mass in a non trivial way, even when each term individually diverges in such limit.  
\end{abstract}

\maketitle

\section{ Introduction }

    The fundamental description of hadronic phenomena in quantum chromodynamics (QCD) presents severe and long-standing problems. The mathematical complexity of the theory, the impossibility of applying perturbation theory in the infrared regime, and the unapproachable need to deal with an infinite number of degrees of freedom have led many researchers to focus on phenomena that take place in a narrow energy range, adopting approximations and constructing effective theories or models valid for specific regimes. Indeed, it is not trivial how to reproduce QCD features of hadrons that occur at different energy scales within the same approach. While at low energy quantum numbers of hadrons seem to be well explained by a small number of constituents, at high energies asymptotic freedom~\cite{gross_ultraviolet_1973,Gross:1974cs,Politzer:1974fr,Politzer:1973fx} of many weakly-interacting particles dominates~\cite{Feynman:1969ej}. 

    The renormalization group procedure for effective particles (RGPEP) was developed to solve bound states in QCD. The method stems from the similarity renormalization group, formulated by G\l azek and Wilson~\cite{glazek_renormalization_1993,glazek_similarity_1997}. It is formulated in the front form of Hamiltonian dynamics~\cite{dirac_forms_1949} and introduces the notion of effective particles in quantum field theory~\cite{glazek_effective_2017}. 
    
    The RGPEP provides a means for constructing a family of equivalent effective Hamiltonians depending on a scale parameter, which allows connecting different energy scales, and selecting the range of interest without truncating significant matrix elements. 
    It introduces the concept of effective particles~\cite{glazek_effective_2017} which are related to canonical or pointlike ones by means of a unitary transformation. 

    The method has been applied to several theories~\cite{Glazek:2014ria,glazek_computation_2020,Glazek:2021vnw,gomez-rocha_asymptotic_2017,glazek_renormalized_2017,Serafin:2018aih}, some of them simple enough to be solved exactly~\cite{Glazek-scalar-2012,Glazek_fermion-20213}. In the case of QCD, only perturbative solutions in powers of the coupling constant have been considered so far in the renormalization-group equation. For instance, the eigenvalue problems for heavy quarkonium and heavy baryons have been derived starting from QCD with only heavy quarks using the second-order Hamiltonian solution to the RGPEP equation. The quark-(anti)quark effective potential for effective quarks and gluons has been obtained and the spectra of hadrons have been computed with unexpected accuracy~\cite{glazek_effective_2017,Serafin:2018aih}. 
    
    Third-order calculations allowed us to analyze the structure of the three-gluon vertex~\cite{glazek_dynamics_2001,gomez-rocha_asymptotic_2015}. 
    From the coefficient of such structure, the running of the coupling constant was defined as its variation with the renormalization group scale $\lambda$. The method turned out to reproduce correctly the property of asymptotic freedom~\cite{glazek_dynamics_2001,gomez-rocha_asymptotic_2015}.
    Moreover, it was shown that the running of the coupling had a similar or identical form to the variation with virtual momenta that is obtained using Feynman diagrams for off-shell Green's functions. 
    However, in Refs.~\cite{glazek_dynamics_2001,gomez-rocha_asymptotic_2015}, three different regularizations were considered, and a certain finite dependence on the choice of the regulating function was encountered, which needs deeper understanding.

    This article concerns the study of a new regularization method for the QCD Hamiltonian within the RGPEP. 
    Namely, we introduce an infinitesimal gluon-mass parameter $m_g$ to regulate singularities that appear for gluons carrying a small longitudinal momentum fraction $x$. In other words, we allow gluons to have a non zero canonical mass. This regularization allows us to circumvent the vacuum problem which appears only in cases $p^+=0$, producing infinity eigenvalues for $m_g\to 0$, a feature which is welcome because no free, colored particles are found in nature.
    
    In particular, we consider again the third-order three-gluon-vertex Hamiltonian in the SU(3) Yang-Mills theory and explore the effect of such a regularization in the asymptotic freedom result. 
    On a similar basis to what was done in~\cite{gomez-rocha_asymptotic_2015}, the running coupling is identified as the coefficient in front of the effective three-gluon-vertex Hamiltonian operator, which is derived by solving the RGPEP equation perturbatively. 
    
    Previous works collect a detailed description of formulas appearing at every step taken in perturbative RGPEP calculations~\cite{glazek_perturbative_2012,gomez-rocha_asymptotic_2015}. In this document, we summarize the main steps and focus our attention on the new element considered here: the new regularization and its consequence in the interaction terms and running coupling. 
    We see clearly how the dependence on the gluon mass cancels once all contributions to the three gluon vertex in the limit $m_g\to 0$ are added, even though separate contributions diverge.
    
    The regularization treated here has been used in other problems addressed within the RGPEP~\cite{Glazek:2021vnw,glazek_massive_2019}. Given the success of such studies and aiming at future high-order solutions to the RGPEP equation in QCD, we apply it here to the most difficult physical problem considered so far using this approach.

    Other methods also include a canonical gluon mass for different considerations. For example, a Faddev-Poppov Lagrangian in the Landau gauge with a tree-level mass term \cite{pelaez_2106_2021} can reproduce the pion decay constant \cite{pelaez_2212_2023} and lattice gluon and ghost correlation functions \cite{tissier_1011_2010} by means of perturbative calculations extended to the infrared regime. 
    This is different from other studies on  dynamical gluon-mass generation which is broadly explored from different approaches~\cite{Cornwall:1981zr,Cornwall:2015lna,Bernard:1982my,Bernard:1981pg,aguilar_2111_2022,aguilar_2211_2023,Aguilar:2016ock,Binosi:2017rwj}, including RGPEP~\cite{glazek_renormalized_2017,Serafin:2018aih,QbarQ-new-reg}.

    The article is organized as follows. Section~\ref{sec:method} presents the main and general steps of the RGPEP and the regularization method employed in this particular work. Section~\ref{sec:vertex} provides the derivation of the three-gluon vertex and the resulting running coupling. Section~\ref{sec:conclusions} concludes the article. An Appendix with extended formulas is available at the end of the document.

\section{ The method }
\label{sec:method}

\subsection{Front form of dynamics}
\label{sec:FFdynamics}

The RGPEP uses the front form of dynamics~\cite{dirac_forms_1949,brodsky_quantum_1998}. In this form, four vectors in Minkowski space-time are denoted by 
$x^\mu=(x^+,x^-,x^\perp)$, where $x^+=x^0+x^3$, $x^-=x^0-x^3$, and $x^\perp=(x^1,x^2)$. We use the Brodsky-Lepage notation~\cite{brodsky_quantum_1998}; the inner product is defined as
\begin{eqnarray}
    a\cdot b 
    \es 
    a^\mu b^\nu g_{\mu\nu} 
    \ = \ \h a^+b^- + \h a^-b^+ - a^\perp  b^\perp \ , 
\end{eqnarray}
where summation over repeated indices is assumed. The ``-" component of the four momentum $p^\mu$ is then
\begin{eqnarray}
    p^- \es { p^{\perp 2} + m^2 \over p^+ } \ ,
\end{eqnarray}
and represents the energy of the particle. 

Our starting point is the front-form Hamiltonian of QCD without quarks. The Hamiltonian density can be derived from the $\cT^{+-}$ component of the energy-momentum tensor $\cT^{\mu\nu}$ associated with the corresponding Lagrangian density. Integration of $\cT^{+-}$ over the hypersurface $x^+=0$ leads to the (classical) front-form Hamiltonian
\begin{eqnarray}
P^- 
\es 
\h \int_{x^+=0} dx^- d^2 x^\perp \cH \ .
\end{eqnarray}

Canonical quantization requires replacing the field $A^\mu$  by the quantum operator $\hat A^\mu$, defined by its Fourier expansion on the surface $x^+=0$,
\begin{eqnarray}
\hat A^\mu 
=
\sum_{\sigma c} \int [k] 
[t^c\varepsilon^\mu_{k\sigma}\hat a_{k\sigma c} e^{-ikx} + 
t^c\varepsilon^{\mu*}_{k\sigma} \hat a_{k\sigma c}^\dagger e^{ikx}]_{x^+=0} \ .
\end{eqnarray}
where $[k]=dk^+d^2k^{\perp}/(16\pi^3 k^+)$. 
For simplicity, we omit hats in operators in the sequel.
Creation and annihilation operators, $a_{k\sigma c}^\dagger$ and $a_{k\sigma c}$, satisfy the canonical commutation relation
\begin{eqnarray}
\left[ a_{k\sigma c}, a_{k'\sigma' c'}^\dagger\right]
\es
k^+ \tilde \delta(k-k')\delta^{\sigma \sigma} \delta^{cc'} \ ,
\end{eqnarray}
where $\sigma$ and $c$ are spin and color indices, respectively. The momentum delta function is given by $\tilde \delta(p)=16\pi^3\delta(p^+)\delta(p^1)\delta(p^2)$, and the polarization four vectors have the components
\begin{eqnarray}
 \varepsilon^\mu_{k \sigma} 
 \es 
 \left(
 \varepsilon^+_{k\sigma}=0, \,
 \varepsilon^-_{k\sigma}=2k^\perp\varepsilon^\perp_\sigma/k^+, \, 
 \varepsilon^\perp_\sigma
 \right) \ .
\end{eqnarray}
 
 The quantum canonical Hamiltonian is obtained after normal ordering of operators, represented by: 
\begin{eqnarray}
\hat P^- 
\es 
\h \int dx^- d^2 x^\perp : \cH (\hat A) : \ .
\end{eqnarray} 
Details of $\hat P^-$ can be found in~\cite{gomez-rocha_asymptotic_2015,glazek_dynamics_2001,glazek_harmonic-oscillator_2004}.

For pure gluonic QCD, only products of fields $A^\mu$ and of its derivatives are involved:
\begin{eqnarray}
\cH 
\es
\cH_{A^2} + \cH_{A^3} + \cH_{A^4} + \cH_{[\partial AA]^2} \ ,
\end{eqnarray}
where $\cH_{A^2}$ is the free Hamiltonian, $\cH_{A^3}$ is the 3-gluon vertex, $\cH_{A^4}$ is the 4-gluon vertex and $\cH_{[\partial AA]^2}$ appears due to the constraint imposed on $A^-$ by the Lagrange equations in the light-cone gauge $A^+=0$ (cf.~Refs.~\cite{gomez-rocha_asymptotic_2015,glazek_dynamics_2001,brodsky_quantum_1998} for details):
\begin{eqnarray}
A^-
\es 
2 {1\over \partial^+}\partial^\perp A^\perp -g{2i\over \partial^{+2}} 
\left[ \partial^+A^\perp, A^\perp \right] \ .
\label{eq:constraint}
\end{eqnarray}

We will use the following notation. 
Hamiltonian terms with one creation and one annihilation operators are written using subscripts $H_{11}$. Likewise, terms with  two creation and one annihilation operators are denoted by $H_{21}$. 
More generally, $H_{i j}$, denotes a term with $i$ creation and $j$ annihilation operators.

The kinetic energy of \textit{free} particles is independent of the coupling constant $g$ 
\begin{eqnarray}
H_{11}
\es 
\h \int_{x^+=0}dx^-d^2x^\perp :\cH_{A^2}:
\nn 
\es 
\sum_1 \int [1] { k_1^{\perp 2} + m_{g}^2 \over k_1^+ } 
a_1^\dagger a_1   \   , 
\label{H11}
\end{eqnarray}
where we have used the abbreviation $[1]=[k_1]$. The subscript 1 in particle operators stands for spin $s_1$, color $c_1$, and momentum $k_1$ of particle 1. The parameter $m_{g}$ is the canonical gluon mass, added to the QCD Hamiltonian. An extended discussion about this parameter will be given in the next section, where the regularization method is explained.

\begin{figure}
    \centering
    \includegraphics[width=0.3\textwidth]{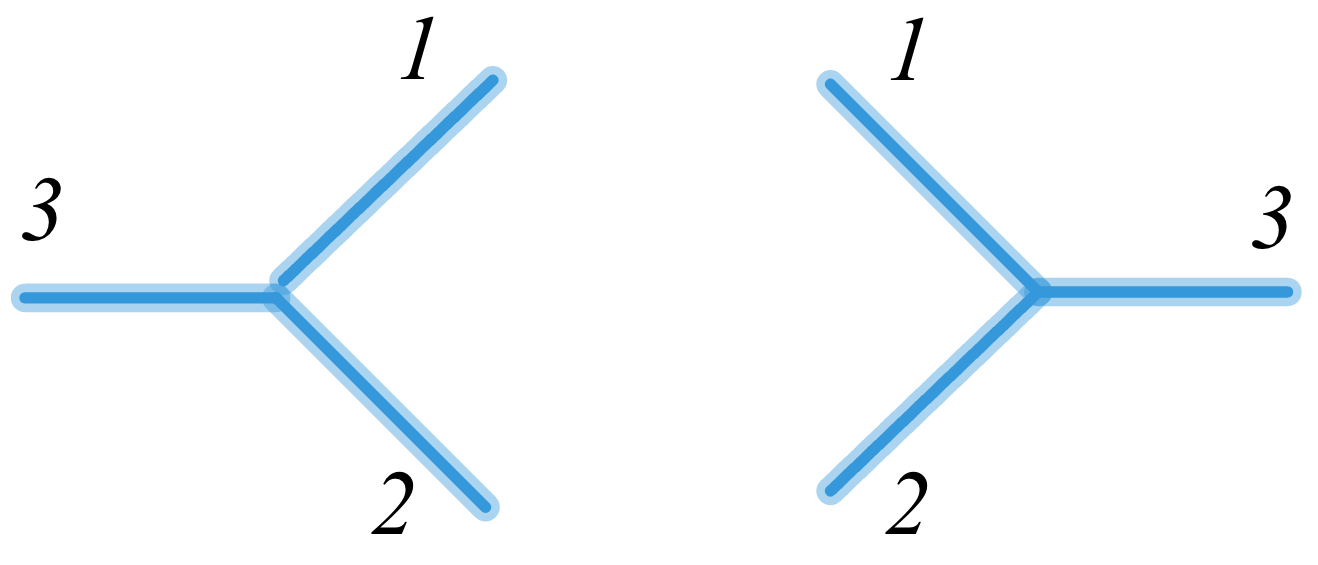}
    \caption{First-order three-gluon vertex diagram. Left: Term with two annihilation and one creation operators, $Y_{123}^*\, a_3^\dagger a_2 a_1$. Right: Term with one annihilation and two creation operators, $Y_{123}a_1^\dagger\, a_2^\dagger  a_3$.}
    \label{fig:1stOrder}
\end{figure}

The three-gluon vertex is proportional to $g$ and has the form
\begin{eqnarray}
H_{21} + H_{12}
\es 
\h \int_{x^+=0} dx^- dx^\perp : \cH_{A^3} :
\label{eq:first-order-vertex}\\
\es
g \sum_{123} \int [123] f_{t_r} \tilde \delta \left( k^\dagger - k\right) Y_{123} a_1^\dagger a_2^\dagger a_3 + h.c,
\nonumber 
\end{eqnarray}
where numbers 1, 2, 3 in the creation and annihilation operators represent all quantum numbers of particles 1, 2, and 3, respectively, and the shortcut notation $[123]=[1][2][3]$ is used. The argument of the Dirac's delta function $\left( k^\dagger - k\right)$ denotes differences of momenta of created particles minus annihilated ones in an interaction respectively. These term is represented in Figure~\ref{fig:1stOrder}.

The momentum carried together by all annihilated or all created particles in a single interaction is called \textit{parent} momentum. 
Conservation of momentum allows us to write $k_1^+=x_1k^+_3$,  $k_2^+=x_2k^+_3$, with $x_1+x_2=1$; and $k_1^\perp = x_1k_3^\perp + \kappa_{12}^\perp$, $k_2^\perp = x_3k_3^\perp - \kappa_{12}^\perp$, so that $\kappa_{12}^\perp = x_2k_1^\perp - x_1 k_2^\perp$.
We use the subscript $i/p$ to denote the momentum of the daughter particle $i$ relative to the parent $p$. For example, in a vertex $a_1^\dagger a_2^\dagger a_3$ the parent momentum is $k_3$ and $k_3^{\perp,+}=k_1^{\perp,+}+k_2^{\perp,+}$. We write $k_1^\perp = x_{1/3}k_3^\perp + \kappa_{1/3}^\perp$,  where $x_{1/3}$ denotes $x_1/x_3$ and $\kappa_{1/3}^\perp=\kappa_{12}^\perp$.

The factor $Y_{123}$ is a polarization function:
\begin{eqnarray}
Y_{123} 
\es 
i f^{c_1 c_2 c_3} 
\left[ 
\varepsilon_1^*\varepsilon_2^*\cdot \varepsilon_3 \kappa
-
\varepsilon_1^*\varepsilon_3 \cdot \varepsilon_2^* \kappa  {1 \over x_ {2/3} }
\right. 
\nn 
&&
\qquad \qquad
\left.
- \ 
\varepsilon_2^*\varepsilon_3\cdot \varepsilon_1^* \kappa {1\over x_{1/3}}
\right] \ .
\label{eq:Y123}
\end{eqnarray}
with $\varepsilon\equiv \varepsilon^\perp$ and $\kappa\equiv \kappa^\perp$. Finally, $f_{t_r}$ in Eq.~(\ref{eq:first-order-vertex}) is a regularization function described in the next section. The subscript $t_r$ is a cutoff parameter. 
For completeness, let us mention that terms $\cH_{A^4}$ and $\cH_{[\partial AA]^2}$ give rise to Hamiltonian terms with four particle operators~\cite{glazek_dynamics_2001,gomez-rocha_asymptotic_2015}.

\subsection{Effective Hamiltonians}

The RGPEP introduces the concept of effective particles. Canonical operators are transformed into effective ones by means of a similarity transformation
\begin{eqnarray}
\label{eq:a_eff}
a_t 
\es 
\cU_t a_0 \cU_t^\dagger  \ ,
\end{eqnarray}
where $\cU_t$ is an anti-Hermitian operator and $t$ the renormalization-group parameter, which can be related to parameters $s$ and $\lambda$ by
\begin{eqnarray}
s = 1/\lambda = t^{1/4} \ .
\label{eq:lambdat}
\end{eqnarray}
The parameter $s$ has units of length and is associated with the \textit{size} of effective particles; the parameter $\lambda$ has units of mass and is associated with the energy \textit{scale}.
While bare operators $a_0^{(\dagger)}$ annihilate (create) bare, pointlike particles, effective particle operators $a^{(\dagger)}_{t}$ annihilate (create) particles of size $s$.

The RGPEP is based on the demand that
\begin{eqnarray}
H_0(a_0) \es H_t(a_t)  \ ,
\label{eq:H0Ht}
\end{eqnarray}
where $H_t$ is the effective Hamiltonian. Denoting $\cH_t = H_t(a_0)$, and differentiating Eq.~(\ref{eq:H0Ht}), one can derive the RGPEP equation 
\begin{eqnarray}
\label{eq:rgpep}
{d \cH_t \over dt}
\es 
\left[ \left[\cH_f , \cH_{Pt}\right], \cH_t\right] \ ,
\end{eqnarray}
with the generator $\cG_t=\left[\cH_f , \cH_{Pt}\right]$ defining the unitary transformation in Eq.~(\ref{eq:a_eff}). The operator $\cH_f$ is the Hamiltonian term that do not contain interactions, Eq.~(\ref{H11}). The operator $\cH_{Pt}$ is identical to $\cH_t$ but it contains a factor equal to $(\sum_i p^+_i)^2$, with $i$ denoting all particles involved in the interaction~\cite{glazek_perturbative_2012}.
The Eq.~(\ref{eq:rgpep}) can be solved order by order using a perturbative expansion of $\cH_t$ in powers of the coupling constant $g$
\begin{eqnarray}
\cH_t 
=
\cH_0 + g\cH_{t\,1} + g^2\cH_{t\,2} + g^3\cH_{t\,3} + g^4\cH_{t\,4} + \dots \quad 
\label{eq:perturb}
\end{eqnarray}

The description of the three-gluon vertex considered here involves powers not larger than third. Following the notation explained above Eq.~(\ref{H11}), this requires terms of the type:
\begin{eqnarray}
\cH_t
\es 
\cH_{11,0,t} + \cH_{21,g,t} + \cH_{12,g,t}
\np
\cH_{11,g^2,t} + \cH_{22,g^2,t} + \cH_{12,g^3,t} + \cH_{21,g^3,t},  \qquad
\label{eq:perturbative-expansion}
\end{eqnarray}
where an additional subscript indicates the power of the coupling constant each term contains. Introducing Eq.~(\ref{eq:perturbative-expansion}) in the RGPEP Eq.~(\ref{eq:rgpep}) yields products of Hamiltonians of different orders. The equation can be then solved order by order (see Appendix~\ref{ap:orderbyorder}). 
From all terms involved in third-order solutions we will focus only on those relevant in the calculation of the three-gluon vertex, namely those with two creation and one annihilation operators and vice versa. 
\begin{figure}
    \centering
    \includegraphics[width=0.26\textwidth]{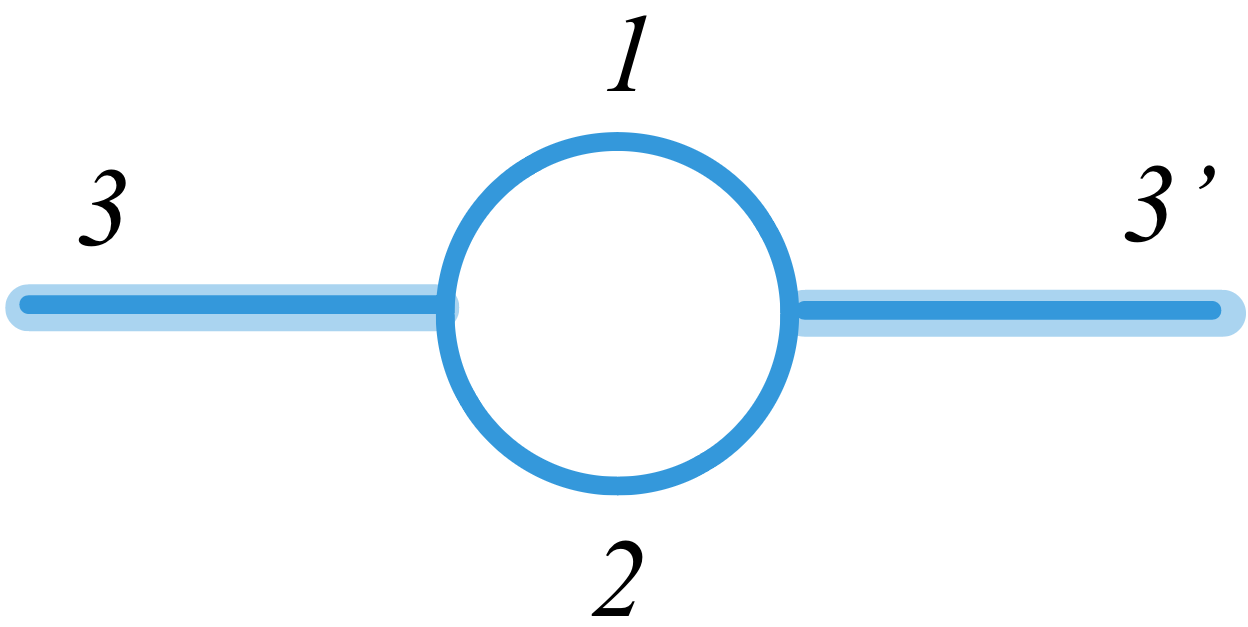}
    \caption{Second-order self-energy diagram resulting from the product of two first-order three-gluon vertices (cf. Figure~\ref{fig:1stOrder}). Thick lines represent effective gluons and thin lines represent creation and annihilation operators that have been eliminated in the normal ordering.}
    \label{fig:2ndOrder}
\end{figure}

\subsection{ Regularization and counterterms }

The canonical Hamiltonian presents ultraviolet and small-$x$ divergences that have to be treated through the RGPEP. Integration of the RGPEP equation yields factors containing exponentials called form factors (cf. Appendix~\ref{ap:orderbyorder})
\begin{eqnarray}
f_{ab,\, t} 
\es 
e^{-ab^2 t} \ , 
\label{eq:ff}
\end{eqnarray}
where $ab:=\cM_a^2-\cM_b^2$ represents difference of invariant masses of particle configurations $a$ and $b$ associated, respectively, with all creation and annihilation operators in a vertex. 
For example, in a vertex $a^\dagger_1 a^\dagger_2 a_3$, cf. Figure~\ref{fig:1stOrder}, the form factor is
\begin{eqnarray}
f_{t \, 12.3} 
\es 
{\exp\left[-t(\cM_{12}^2-\cM_{3}^2)^2\right]} \ ,
\end{eqnarray}
with invariant masses defined as $\cM_{i...j}^2 = \left(p_{i}+...+p_{j}\right)^2$. For single gluons $\cM_{i}^2=m_{g}^2$ and for pairs $\cM_{ij}^2
=\frac{\kappa_{ij}^2+m_{g}^2}{x_{i/p}x_{j/p}}$, where $x_{i/p} = x_{i} / x_{P}$, and $x_p=x_i+x_j$. Note that, in distinction to the previous work~\cite{gomez-rocha_asymptotic_2015}, $\cM_3^2$ is not zero and thus, form factors change from 
\begin{equation}
  f_{t\, ij.l}  =   \exp\left(-t\frac{\kappa_{ij}^4}{x_{i}^2x_{j}^2}\right) \ ,  \qquad 
\end{equation}
to
\begin{equation}
     f_{t \, ij.l}  =   \exp\left[-t\left(\frac{\kappa_{ij}^2+m_{g}^2}{x_{i}x_{j}}-m_{g}^2\right)^2\right]  \ .  
\end{equation}
These factors remove all ultraviolet divergences present in a single interaction term.

On the other hand, to tackle the more subtle small-$x$ divergences appearing in loop integrals we introduce a regulating function depending on the small cutoff $t_{r}$ in each vertex 
\begin{eqnarray}
f_{t_r \, 12.3} 
\es 
{\exp\left[-t_r(\cM_{12}^2-\cM_{3}^2)^2\right]} \ .
\end{eqnarray}
The behavior of form factors with and without a gluon mass is different around $\kappa_{ij} = 0$ and $x_{i},x_{j} = 0$, precisely the region in which loop integrals diverge. 
The gluon mass and $t_{r}$ remove all small-$x$ divergences. 

The result in the limit $t_r\to 0$ and $m_g\to 0$ must be then analyzed and any possible remaining divergence should be canceled by counterterms.

Thus, the counterterms provide the initial condition of the differential equation~(\ref{eq:rgpep}), and is such that in the limit in which the cutoffs are lifted, the canonical Hamiltonian must be
recovered and any cutoff dependence must be removed.
Therefore, the initial Hamiltonian consists of the (regularized) canonical Hamiltonian plus counterterms. 

To illustrate the procedure in a simple case, consider the self-energy contribution resulting from the product of two first-order vertices (cf.~Fig.~\ref{fig:2ndOrder}):
 \begin{eqnarray}
\hat{\mu}_{t,ab}^{2}
\es 
2 \sum_{1}\int\left[1\right]
\frac{1}{p_{1}^{+}}
\nt 
\left\{ \int\left[x\kappa\right]\frac{\kappa^{2}P\left(x\right)}{2x\left(1-x\right)}\frac{f_{t+t_{r}}^{2}-f_{t_{r}}^{2}}{\mathcal{M}^{2}-m_{g}^{2}}\right\} a_{1}^{\dagger}a_{1}+\hat{\mu}_{0,ab}^2 \ ,
\label{eq:mut}
\nn
\end{eqnarray}
where $P(x)$ is the Altarelli-Parisi gluon splitting function~\cite{Altarelli:1977zs}.

Ultraviolet divergences are regularized by form factors $f_{t}$ that suppress interactions with changes of invariant masses greater than about $\lambda=t^{-1/4}$. 
But terms that do not contain that function may diverge and a counterterm is required to cancel the cutoff dependence introduced by regulating functions.  In Eq.~(\ref{eq:mut}), the counterterm is represented by $\hat{\mu}_{0,ab}^2$.

There are also small-\textit{x} divergences that cannot be regulated by the RGPEP and that arise from \textit{zero modes}. 
The presence of a gluon mass avoids the appearance of such infinities. 
On the other hand, the divergence of the self-energy due to small-\textit{x} singularities is not a problem because an isolated gluon is not a physical state, and thus its mass is not an observable.

The counterterm is
\begin{eqnarray}
&&\hat{\mu}_{0,ab}^2
=
2
 \sum_{1}\int\left[1\right]
\frac{1}{p_{1}^{+}}\left\{ \int\left[x\kappa\right]\frac{\kappa^{2}P\left(x\right)}{2x\left(1-x\right)}\frac{f_{t_{r}}^{2}}{\mathcal{M}^{2}-m_{g}^{2}}
\right. 
\nn
&&
\left.
\qquad \qquad \qquad \qquad + \ \mu_{0,m_{g}}^{2}\right\} a_{1}^{\dagger}a_{1},
\end{eqnarray}
and the self-energy term can be finally written as
\begin{eqnarray}
\hat{\mu}_{t,ab}^{2}
\es
2
\sum_{1}\int\left[1\right]
\frac{\mu_{t}^{2}}{p_{1}^{+}}a_{1}^{\dagger}a_{1} 
\nn
\es 
2
\sum_{1}\int\left[1\right]
\frac{1}{p_{1}^{+}}
\left[\int\left[x\kappa\right]
\frac{\kappa^{2}P\left(x\right)}{2x\left(1-x\right)}\frac{f_{t+t_{r}}^{2}}{\mathcal{M}^{2}-m_{g}^{2}}
\right. 
\nn
&& \quad + \ 
\left.
\mu_{0,m_{g}}^{2}\right]a_{1}^{\dagger}a_{1} \ .
\label{mut}
\end{eqnarray}
The remaining dependence on $t_{r}$ vanishes because this parameter appears in addition to $t$ in form factors~\cite{glazek_computation_2020}. In contrast, note that the gluon mass is still necessary to regulate small-\textit{x} divergences. Finally, $\mu_{0,m_{g}}^{2}$ is an ultraviolet finite remnant that cannot be fixed by the RGPEP procedure alone and, as the subscript $m_{g}$ indicates, it can be divergent in the small-\textit{x} limit and thus used to cancel the divergence on $\mu_{t}^2$, in case we need it to be finite~\cite{wilson_nonperturbative_1994}. 

Eq.~(\ref{mut}) can be integrated analytically in the limit of small mass $m_{g}$ using the procedure described in Appendix~\ref{integration}. After the integration, one can identify divergent and finite terms in the resulting expression
\begin{equation}
\mu_{t}^2=-\sqrt{\frac{\pi}{2}}\frac{1}{4\sqrt{t}}\left[\log\left(tm_{g}^{4}\right)+\log\left(8e^{\gamma_{E}}\right)+\frac{23}{6}\right] + \mu_{0,m_{g}}^{2},
\label{self-energy result}
\end{equation}
where $\gamma_{E}$ is the Euler-Mascheroni constant. Note how clearly $\mu_t^2\to \infty$ when $m_g\to 0$.


\begin{figure}
    \centering
    \includegraphics[scale = 0.9]{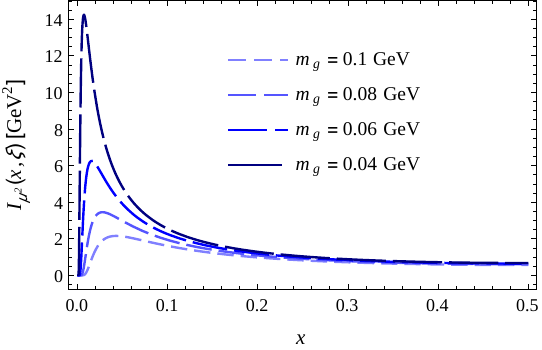}
    \caption{Integrand of Eq.~(\ref{mut}) after numerical integration over $\kappa^\perp$ for different values of the parameter $m_{g}$.
    The gluon mass acts as a regulator: the smaller its value, the more is the tendency to infinity of the integrand for $x$ close to 0. The behavior is symmetric for $0.5<x<1$.}
    \label{fig:my_label}
\end{figure}

\section{ Three-gluon vertex and running coupling }
\label{sec:vertex}

\begin{figure}
    \centering
    \includegraphics[width=0.3\textwidth]{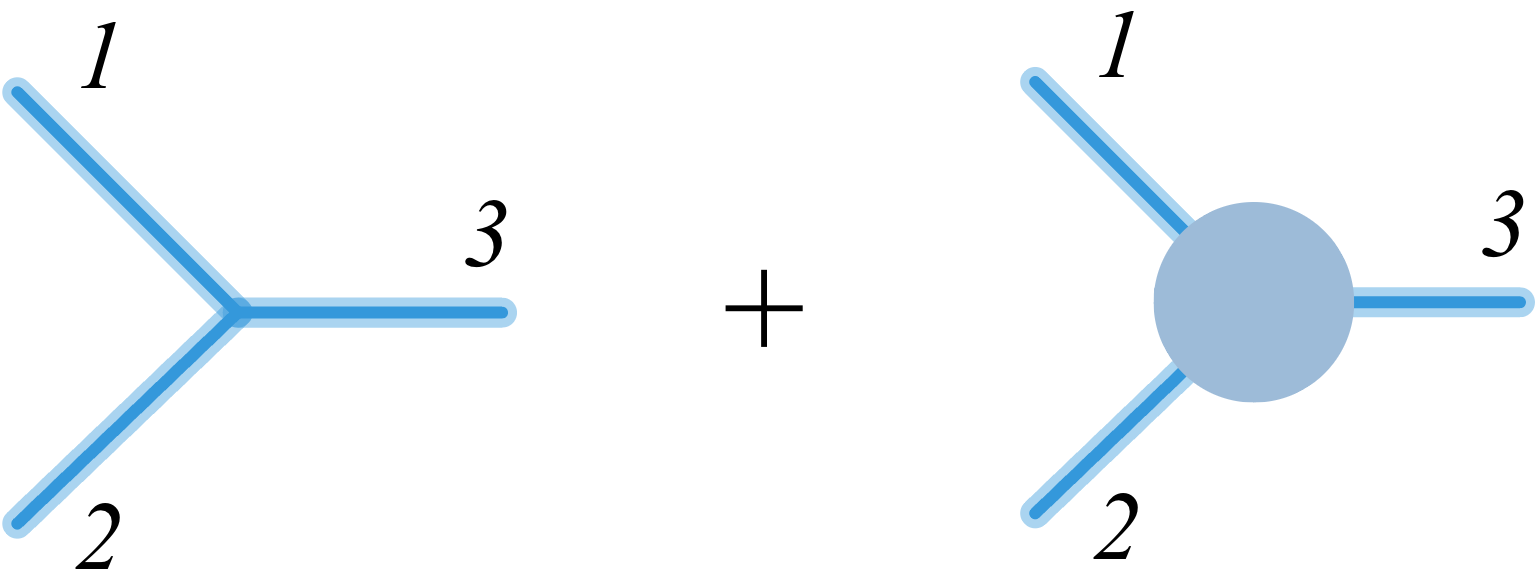}
    \caption{Illustration of the three-gluon vertex as a sum of structures with one annihilation and two creation of effective gluons. The first term is proportional to $g$ and the second one is proportional to $g^3$. The latter is a sum of different structures represented in Figure~\ref{fig:3rdOrder}. }
    \label{fig:3gluonvertex}
\end{figure}

From the expansion of Eq.~(\ref{eq:perturb})-(\ref{eq:perturbative-expansion}) we identify the three-gluon-vertex structure in those terms that annihilate (create) one effective gluon and create (annihilate) two ones. Indeed, the expression (cf. Figure~\ref{fig:3gluonvertex})
\begin{widetext}
    
\begin{equation}
\cV_{21t}
=
g\cY_{21t}+g^3\cK_{21t}
=
\sum_{123}\int[123]\tilde{\delta}\left(p^{\dagger}-p\right)f_{t,ab}
\left\lbrace g
\tilde{\cY}_{21t}
+ g^3\left(
\tilde{\cK}_{21t}
+
\tilde{\cK}_{210}
\right)\right\rbrace a_{t1}^{\dagger}a_{t2}^{\dagger}a_{t3} \ ,
\label{third-order-vertex}
\end{equation}

\end{widetext}
has first- and third-order terms.

\begin{widetext}

\begin{figure}
    \centering
    \includegraphics[width=0.8\textwidth]{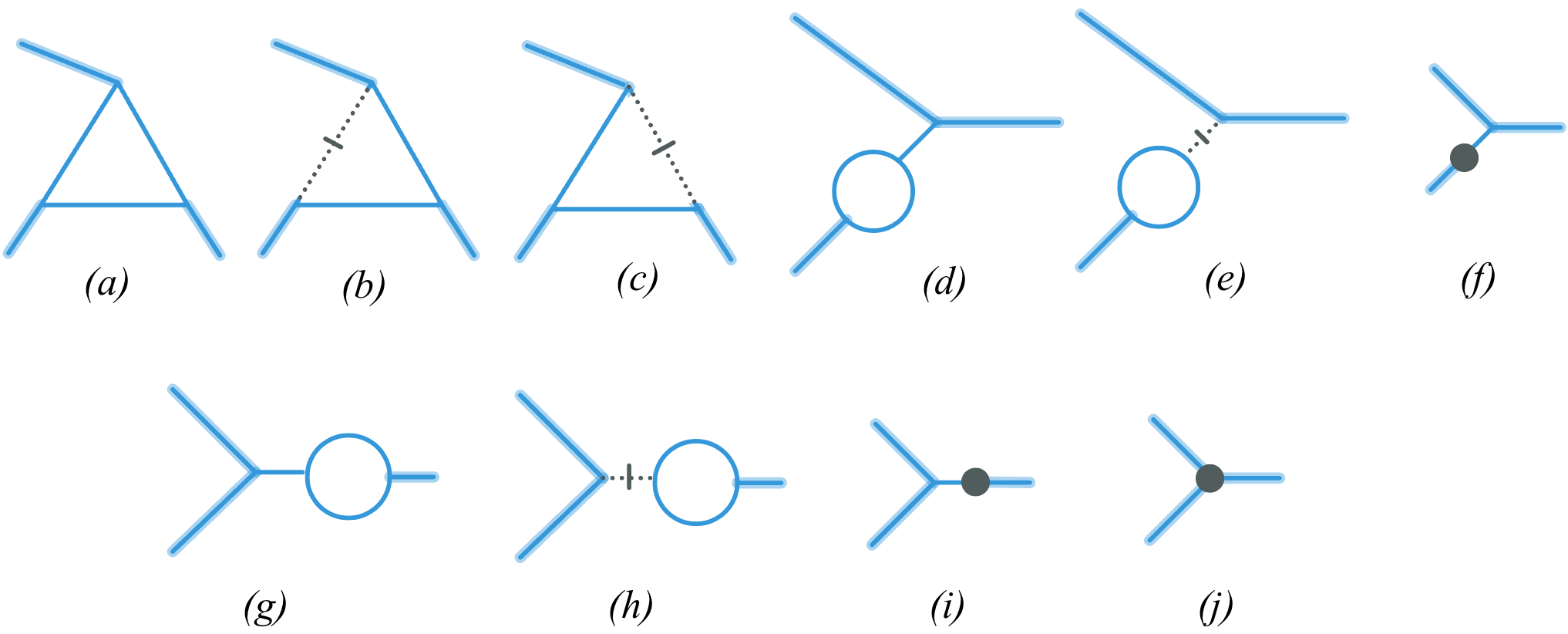}
    \caption{Third-order diagrams resulting from products of first-order terms [diagrams \textit{(a)}, \textit{(d)}, \textit{(g)}], and from products of first-order times second-order terms [diagrams \textit{(b)}, \textit{(c)}, \textit{(e)}, \textit{(f)}, \textit{(h)}, \textit{(i)}]. The thick dots in \textit{(f)} and \textit{(i)} represent self-energy counterterms. Diagram \textit{(j)} is the third-order counterterm that removes loop divergences arising at this order when the cutoff is lifted. Thicker lines represent effective creation/annihilation operators. Thin internal lines represent operators that are eliminated in the normal ordering. Dotted lines with a transverse dash indicate instantaneous interactions present in the canonical Hamiltonian due to the constraint of Eq.~(\ref{eq:constraint}).}
    \label{fig:3rdOrder}
\end{figure}

\end{widetext}

The first-order term is given by Eq.~(\ref{eq:first-order-vertex}) but including now the form factor $f_t$; and with the additional difference that particle operators are of type $t$ and have a width $s$.
The third-order term is a sum of products of interactions that result from introducing Eq.~(\ref{eq:perturb}) in Eq.~(\ref{eq:rgpep}).
We can distinguish nine different types of contributions depending on the operator product contained in them.
They are represented diagrammatically in  Figure~\ref{fig:3rdOrder}.

In terms of these diagrams the third order component in Eq.~(\ref{third-order-vertex}) is written as
\begin{equation}
\tilde{\cK}_{21t}=\frac{\sum_{n}\gamma\left(n\right)}{2\cdot16\pi^{3}} \ ,    
\label{gamma-def}
\end{equation}
where $n$ ranges from $a$ to $i$, and $j$ corresponds to the third-order counterterm. 
Similarly to the self-energy Hamiltonian term, $\mu_{t}^{2}$ in Eq.~(\ref{mut}), each $\gamma\left(n\right)$ involves three-dimensional loop integrals over the longitudinal momentum fractions $x$ and relative transverse momenta $\kappa^\perp$ of the internal virtual particles. 
A detailed expression for the integrand of each of these structures can be found in~\cite{gomez-rocha_asymptotic_2015}, with the unique difference that in this work  all invariant masses contain the canonical gluon mass $m_{g}$, which makes the integration procedure much more involved.

By inspection of the three-gluon vertex, the effective coupling depending on the RGPEP parameter \textit{t} can be defined~\cite{glazek_dynamics_2001,gomez-rocha_asymptotic_2015}.
The running coupling is identified as the coefficient in front of the canonical color, spin and momentum dependent factor $Y_{123}\left(x_1,\kappa_{12},\sigma\right)$ in the limit $\kappa_{12}\approx 0$ for some value of $x_{1}$~\cite{glazek_dynamics_2001,gomez-rocha_asymptotic_2015} because
in this limit, the dressed vertex function factorizes in a way proportional to the bare vertex Eq.~(\ref{eq:first-order-vertex})
\begin{eqnarray}
\tilde{\mathcal{K}}_{21t}\left(x_{1},\kappa_{12},\sigma\right)
\approx
c_{t}\left(x_{1}\right)Y_{123}\left(x_{1},\kappa_{12},\sigma\right)
, 
\label{eq:Y-factorization}
\end{eqnarray}
where the dependence on $\kappa_{12}$ is contained in $Y_{123}$.

The coupling constant $g_t$ must be set to a specific value $g_0$ at some arbitrary scale $t=t_0$ in agreement with experiments describing data using the effective Hamiltonian at $t=t_0$. The running coupling can be expressed in terms of the scale of reference.

Only the third-order counterterm remains to be discussed. Third-order contributions $\gamma\left(n\right)$ become independent of $t$ in the ultraviolet limit $\kappa\rightarrow\infty$, and only form factors with vanishingly small $t_{r}$ remain.
Thus, the difference between two scales $\gamma\left(n\right)_{t}-\gamma\left(n\right)_{t_{0}}$ is ultraviolet finite regardless the values of $t$ and $t_{0}$ and the ultraviolet-divergent part of the counterterm is taken to be 
\begin{equation}
     \tilde{\mathcal{K}}_{210}|_{\text{UV}}=-\sum_{n=a}^{i}\gamma\left(n\right)_{t_{0}},
\end{equation}
while the finite part $\left(\text{in the limit } \kappa\rightarrow\infty\right)$ remains undetermined. For convenience we factorize the structure $Y_{123}$:
\begin{equation}
     \tilde{\mathcal{K}}_{210}|_{\text{UV-finite}}=\tilde c_{0}\left(x_{1},\kappa_{12},\sigma\right)Y_{123}\left(x_{1},\kappa_{12},\sigma\right),
\end{equation}
where in general $\tilde c_{0}$ depends on all variables except the RGPEP scale \textit{t}, and should be further constrained by experimental considerations \cite{glazek_boost-invariant_1999,glazek_dynamics_2001,gomez-rocha_asymptotic_2015}.

Taking this into account, the three-gluon vertex at first order in $\kappa_{12}^\perp$ has the structure

\begin{widetext}
\begin{align}
    \tilde{\cV}_{21t} & =\,g\tilde{\cY}_{21t}+g^3\left(\tilde{\cK}_{21t}+\tilde{\cK}_{210}\right)
    =\left[g+g^{3}\left(c_{t}\left(x_{1}\right)-c_{t_{0}}\left(x_{1}\right)+\tilde c_{0}\left(x_{1},\kappa_{12},\sigma\right)\right)\right] Y_{123}
     \ + \mathcal{O}\left(\kappa_{12}^{2},g^{4}\right) \ .
\end{align}

The running coupling is identified as the coefficient of  $Y_{123}$
\begin{eqnarray}
    g_{t} \equiv   g
    \ + \ 
    g^{3}\left(c_{t}\left(x_{0}\right)-c_{t_{0}}\left(x_{0}\right)
    + \lim_{\kappa_{12}\rightarrow0}\tilde c_{0}\left(x_{0},\kappa_{12},\sigma\right)\right) + \mathcal{O}\left(g^{4}\right) \ ,
\end{eqnarray}
setting $g_{t}$ to be $g_{0}$ at the scale $t_{0}$ we are left with $g_{0} = g+g^{3}\lim_{\kappa_{12}\rightarrow0}\tilde c_{0}\left(x_{0},\kappa_{12},\sigma\right)$:
\begin{equation}
    g_{t} =  g_{0}+g_{0}^{3}\left(c_{t}\left(x_{0}\right)-c_{t_{0}}\left(x_{0}\right)\right) + \mathcal{O}\left(g_{0}^{4}\right) \ .
\end{equation}

\end{widetext}

\subsection{Term \textit{a}}
The triangle term $(a)$ gathers products of three first-order vertices Eq.~(\ref{H11}):

\begin{eqnarray}
&&\left(\mathcal{\tilde K}_{21t}\left(x_{1},\kappa_{12},\sigma\right)
+
\mathcal{\tilde K}_{210}\left(x_{1},\kappa_{12},\sigma\right)\right)|_{a}
\nn
&& \ = \
\cY_{0,ax}\cY_{0,xy}\cY_{0,yb}\left(
C_{t,axyb}
-
C_{t_{0},axyb}\right),
\nn
\end{eqnarray}
%
%
where the contribution of the third-order counterterm is included. $\cY_{0,\alpha\beta}$ are first-order interaction vertices from the initial Hamiltonian, with subscripts denoting particle configurations before and after each interaction. 
The RGPEP factor $C_{t,axyb}$ results from solving the RGPEP equations and it is made of functons of invariant masses and form factors (see Appendix~\ref{ap:orderbyorder}, Figure~\ref{fig:configurations}, and Ref.~\cite{glazek_perturbative_2012,gomez-rocha_asymptotic_2015}).

The product of first-order vertices Eq.~(\ref{eq:first-order-vertex}) that leads to diagram ($a$) is
\begin{widetext}

\begin{equation}
\cY_{0,ax}\cY_{0,xy}\cY_{0,yb}
=
8
\sum_{123}\int[123]\int\frac{dx\,d^{2}\kappa}{16\pi^{3}}\frac{f_{t_{r},ax}f_{t_{r},xy}\,f_{t_{r},yb}}{p_{3}^{+2}}\,\frac{\sum_{678}Y_{682}^{*}\,Y_{167}\,Y_{783}}{(x-x_{1})(1-x)x}\,\bar{\delta}(p_{7}+p_{8}-p_{3})\,a_{1}^{\dagger}a_{2}^{\dagger}a_{3},
\label{first-order-multiplication}
\end{equation}
where the numbers 6, 7, 8 label intermediate gluons. The contribution of this diagram is:
\begin{align}
\left(\gamma_{t}(a)-\gamma_{t_{0}}(a)\right) & =4iN_{c}Y_{123}f^{c_{1}c_{2}c_{3}}\int_{x_{1}}^{1}\frac{dx\,\epsilon^{ijk}(a)}{\left(x-x_{1}\right)\left(1-x\right)x}\int d^{2}\kappa\,f_{t_{r},ax}f_{t_{r},xy}f_{t_{r},yb}\kappa_{68}^{i}\,\kappa_{16}^{j}\,\kappa^{k}\,\frac{C_{t,axyb}-C_{t_{0},axyb}}{p_{3}^{+2}}\nn
& +\left(1\leftrightarrow2\right),
\label{gamma_a_integral}
\end{align}

\end{widetext}
where 
$\epsilon^{ijk}(a)$ is a structure that depends on the relative transverse momentum of external gluons $\kappa_{12}^{\perp}$, and emerges from the product of polarizations: 
\begin{equation}
\sum_{678}Y_{682}^{*}\,Y_{167}\,Y_{783}=\frac{N_{c}}{2}\,if^{c_{1}c_{2}c_{3}}\,\kappa_{68}^{i}\kappa_{16}^{j}\kappa^{k}\,\epsilon^{ijk}(a).
\end{equation}
In order to evaluate Eq.~(\ref{gamma_a_integral}) we simplify the integrand proceeding
    as described in Appendix~\ref{integration}. The  integration over $x$ is divided into regions called \textbf{I}, \textbf{II}, and \textbf{III}: $\left(x_{1},x_{1}+\bar{m}_{g}\right)$, $\left(x_{1}+\bar{m}_{g},1-\bar{m}_{g}\right)$ and $\left(1-\bar{m}_{g},\bar{m}_{g}\right)$, respectively, where $\bar{m}_{g}$ is dimensionless and depends on $m_{g}$. The integral over $x$ diverges in regions~\textbf{I} and~\textbf{III} but is finite in region~\textbf{II}.
\begin{itemize}

    \item \textbf{Regions II}:
    Since there is no divergence in this interval, limits $m_{g} \to 0$ and $t_{r} \to 0$ can be applied and the result is analogous to the one obtained in~\cite{gomez-rocha_asymptotic_2015}.  
    \item \textbf{Regions I and III}:
    The first and third intervals are evaluated separately by expanding the integrals over $x \approx x_{1}$ and $x \approx 1$, respectively, and keeping only the dominant powers. Then, an expansion around $\kappa_{12} \approx 0$ is applied, yielding an expression proportional to the structure $Y_{123}$, cf. Eq.~(\ref{eq:Y-factorization}) with integrals that can be evaluated analytically.

\end{itemize}
   
The total contribution of diagram ($a$) is obtained after adding the result from each interval. A logarithmic dependence on the gluon mass parameter $m_{g}$ remains: 
\begin{eqnarray}
    {\gamma_{t}\left(a\right)-\gamma_{t_{0}}\left(a\right)}
    & \to &
    N_{c}Y_{123}\pi\log\left(\frac{t}{t_{0}}\right)\left[-\frac{11}{3}+\frac{1}{6}h_{a}\left(x_{1}\right)\right]
    \nm
    \frac{16\pi N_{c}Y_{123}}{x_1x_2 } \frac{\bar{t}-\bar{t}_{0}}{x_{1}^2+x_{2}^2}
    \frac{\bar{m}_g}{\bar{t}_{r}} \  ,    
    \label{terma}
\end{eqnarray}
where 
\begin{eqnarray}
\frac{1}{6}h_{a}\left(x_{1}\right)
\es 
-3\log\left(\bar m_g^{4}\sqrt{\bar{t}\bar{t}_{1}}e^{\gamma_{E}}\right) - 5  -\log\left(2\right) 
\np
\frac{1-x_{1}^2x_{2}^2}{\left(1+x_{1}^2\right)\left(1+x_{2}^2\right)}
\nm 
\frac{2}{1-x_{2}^{2}}\log\left(\frac{1+x_{2}^{2}}{x_{1}x_{2}}\right)
-\frac{2}{1-x_{1}^{2}}\log\left(\frac{1+x_{1}^{2}}{x_{1}x_{2}}\right)
\np 
\left(1-\frac{1}{1-x_{1}^{2}}-\frac{1}{1-x_{2}^{2}}\right)
\nt 
\log\left[\frac{\left(x_{1}^{2}+x_{2}^{2}\right)x_{1}^{2}x_{2}^{2}}{2\left(1+x_{2}^{2}\right)\left(1+x_{1}^{2}\right)}\right], 
\end{eqnarray}
being $\gamma_{E}$ the Euler-Mascheroni constant.

\begin{figure}
    \centering
    \includegraphics[scale = 0.9]{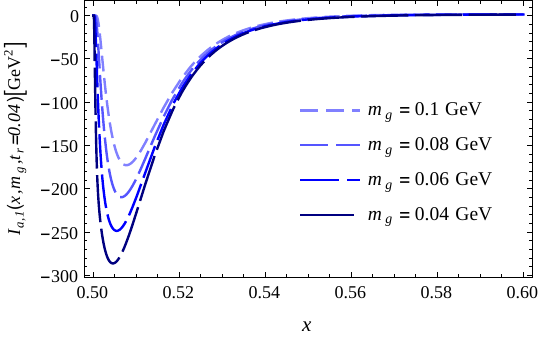}
    
    \includegraphics[scale = 0.9]{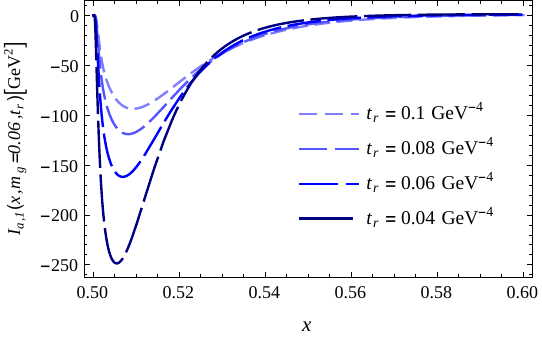}
    \caption{Result of integration over $\kappa^{\perp}$ for region $\mathbf{I}$ of term $\gamma\left(a\right)$ (see Appendix~\ref{integration}). On the upper panel $t_{r}$ is fixed to $0.04 \, \text{GeV}^{-4}$ while on the bottom panel $m_{g}$ is fixed to $0.06 \,\text{GeV}$. In  both figures $t = 2 \,\text{GeV}^{-4}$, $t_{0} = 1 \,\text{GeV}^{-4}$ and $x_{1} = 0.5$. Divergences occur both when $t_{r} \rightarrow 0$ or $m_{g}\rightarrow 0 $.}
    \label{fig:terma_mgcte}
\end{figure}

\subsection{Terms \textit{(b)} to \textit{(i)}}
Terms $(b)$ and $(c)$ are products of the first-order vertex and the second-order canonical interaction. For the term~$(b)$ we have
\begin{equation}
\gamma_{t}\left(b\right)-\gamma_{t_{0}}\left(b\right)
\ \to \  
\frac{16\pi N_{c}Y_{123}}{x_{0}x_{2}}\frac{\left(\bar{t}-\bar{t}_{0}\right)}{x_{1}^{2}+x_{2}^{2}}\frac{\bar m_g }{\bar{t}_{r}}.
\label{termb}
\end{equation}
This contribution exactly cancels the term proportional to $\bar{t}-\bar{t}_{0}$ in Eq.~(\ref{terma}).

The contribution of $(c)$ turns out to be negligible in the limits $m_g \to 0$ and $t_r\to 0$.

Terms \textit{(e)} and \textit{(h)} also result from the product of the same kind of interactions as $(b)$ and $(c)$. They do not contribute to the running coupling due to the absence of linear terms in $\kappa_{12}$ that could give rise to the canonical polarization structure $Y_{123}$ of Eq.~(\ref{eq:first-order-vertex}).

Terms $(d)$ and $(f)$ are also obtained from products of first-order interactions, two of them forming a second-order self-energy contribution $\hat{\mu}_{t}$. Thus, diagrams $(d)$ and $(f)$ come from the first and second terms in Eq.~(\ref{mut}), respectively, the latter containing the counterterm. Their sum gives the following result
\begin{eqnarray}
    &&
    \gamma_{t}\left(d+f\right)-\gamma_{t_{0}}\left(d+f\right)
    \nn
    && \to \ 
    \pi N_{c}Y_{123}\log\left(\frac{t}{t_{0}}\right)\left[\frac{11}{3}+\frac{1}{6}h_{d+f}\left(x_{1}\right)\right] \ ,
\label{termdf}
\end{eqnarray}
where 
\begin{align}
    &\frac{1}{6}h_{d+f}\left(x_{1}\right) \ = \ 2\log\left(e^{\gamma_{E}}\bar m_g^{4}\sqrt{\bar{t}\bar{t}_{0}}\right)+2\log{2}+4 \nonumber \\
    &- 2\frac{x_{2}^{2}}{1-x_{2}^{2}}\log\left(\frac{1+x_{2}^{2}}{2x_{2}^{2}}\right)-2\frac{x_{1}^{2}}{1-x_{1}^{2}}\log\left(\frac{1+x_{1}^{2}}{2x_{1}^{2}}\right) \  .
\label{hdf}
\end{align}

Terms $(g)$ and $(i)$ are analogous, but the loop is located in another leg of the diagram
\begin{eqnarray}
&&\gamma_{t}\left(g+i\right)-\gamma_{t_{0}}\left(g+i\right)
\nn 
&& \to \
\frac{\pi N_{c}Y_{123}}{6}\log\left(\frac{t}{t_{0}}\right)\left[11+h_{g+i}\left(x_{1}\right)\right] \ ,
\label{termgi}
\end{eqnarray}
with
\begin{equation}
\frac{1}{6}h_{g+i}\left(x_{1}\right) \ = \ \log\left(e^{\gamma_{E}}\bar m_g^4\sqrt{\bar{t}\bar{t_{0}}}\right)+\log{2}+1\ .
\label{hgi}
\end{equation}

The final expression of the running coupling is given by Eqs.~(\ref{terma})-(\ref{hgi}) 
\begin{equation}
    g_{\lambda} = g_{0}-N_{c}\frac{g_{0}^{3}}{48\pi^{2}}\log\left(\frac{\lambda}{\lambda_{0}}\right)\left[11+h\left(x_{1}\right)\right],
\label{Rcfinal}
\end{equation}
with $\lambda = 1/\sqrt[4]{t}$, and
\begin{align}
\begin{split}
&h\left(x_{1}\right) =-6\left\{\frac{1+x_{2}^{2}}{1-x_{2}^{2}}\log\left[\frac{\left(1+x_{2}^{2}\right)^{2}}{x_{2}^{2}}\right]\right.\\
&+\frac{1+x_{1}^{2}}{1-x_{1}^{2}}\log\left[\frac{\left(1+x_{1}^{2}\right)^{2}}{x_{1}^{2}}\right]-\frac{1-x_{1}^2x_{2}^2}{\left(1+x_{1}^2\right)\left(1+x_{2}^2\right)}\\
&\left.+\left(1-\frac{1}{1-x_{1}^{2}}-\frac{1}{1-x_{2}^{2}}\right)\log\left[\frac{8\left(1+x_{2}^{2}\right)\left(1+x_{1}^{2}\right)}{\left(x_{1}^{2}+x_{2}^{2}\right)}\right]\right\}.
\end{split}\label{eq:h}
\end{align}

As one can see, the obtained running coupling does not contain any explicit dependence on cutoff $t_r$ or on the gluon mass $m_g$. All possible dependencies have been canceled after adding the contributions. Thus the final result is finite and independent of $m_g$ in the limit $m_g\to 0$, even when the contributions diverge individually in this limit, cf.~Figure~\ref{fig:terma_mgcte}. 

\begin{widetext}    

\begin{figure}
    \centering
    \includegraphics[width=0.55\textwidth]{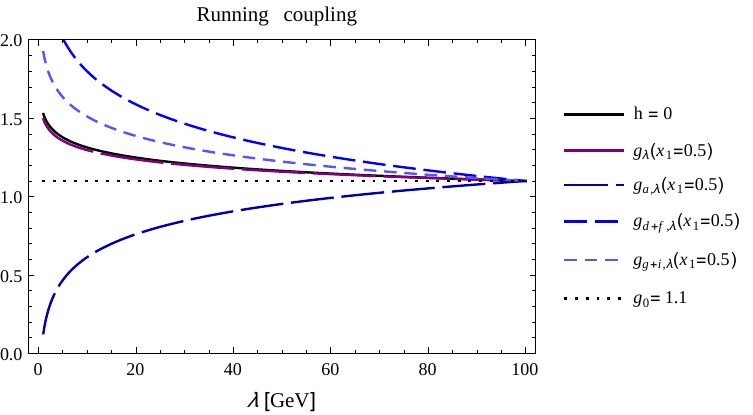}    
    \includegraphics[width=0.4\textwidth]{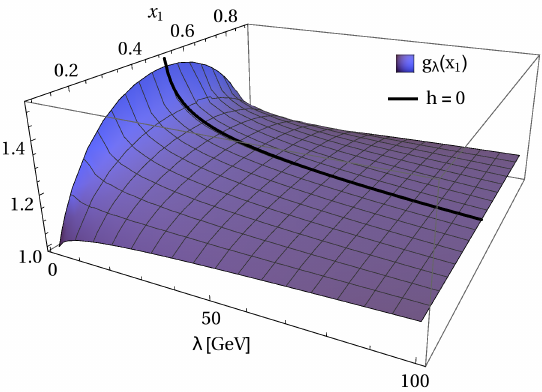}
    \caption{Left: The resulting running coupling Eq.~(\ref{Rcfinal}) varying with $\lambda$ for $x_1=0.5$ (purple). It results from the sum of the contributions coming from the different diagrams (blue dashed lines). It appears very close to the running coupling result corresponding to $h(x_1)=0$, bold black line (cf. Ref.~\cite{glazek_dynamics_2001,gomez-rocha_asymptotic_2015}).
    Right: The function $g_\lambda(x_1)$ shows a dependence on $x_1$ due to the function $h(x_1)$ appearing in Eq.~(\ref{eq:h}), differing from the case $h(x_1)=0$ as $x_1$ approaches the ends $x_1=1$ and $x_1=0$. The difference is more pronounced for lower values of $\lambda$.}
    \label{fig:running_coupling}
\end{figure}
    
\end{widetext}

\subsection{ Analysis of results }
\label{sec:analysis}

It is interesting to observe the results shown in Figure~\ref{fig:running_coupling}. The 2D plot shows separate contributions to the running coupling (for $x_1=0.5$) coming from different diagrams of Figure~\ref{fig:3rdOrder}. Contributions $(d)+(f)$ and $(g)+(i)$ are decreasing and therefore asymptotically free, whereas contribution $(a)$ is increasing. All three contributions balance out and their sum is extremely close to the value of the running coupling obtained from standard calculations using Feynman diagrams (bold black line). The difference is due to the appearance of function $h(x_1)$ in Eq.~(\ref{Rcfinal}), which for the considered value $h(0.5)=-0.863$. This small difference does not even reach a maximum of 2\% in the treated range of scale $\lambda$. 

Such a difference grows, however, for values of $x_1$ that distance from the central one $x_1=0.5$, reaching a maximum at $x_1=0$ and $x_1=1$, as it is plotted in the curved surface of the 3D Figure~\ref{fig:running_coupling}. It seems reasonable to define the running coupling at  $x_1=0.5$. Yet, a need for a more adequate definition of the running of the coupling and including higher-order Hamiltonians than third is not discarded.

The difference in sign in the variation of the coupling provided by the different pieces represented in Figure~\ref{fig:3rdOrder} resembles the difference in sign of terms in the effective potential obtained for quarks in the bound state problem~\cite{glazek_effective_2017,QbarQ-new-reg}. In the bound-state equation, self-energy interactions are repulsive, while the exchange of gluons between quarks is attractive. Hence, attractive and repulsive forces combine to form bound states, canceling divergences that appear in them when they are taken separately. 
It is not surprising then, that Hamiltonian terms combine analogously in the three-gluon vertex too. Indeed, terms $(d)+(f)$ and $(g)+(i)$ are self-energy Hamiltonians multiplied by a three-gluon vertex, whereas term $(a)$ can be seen as a gluon exchange term that is contracted with a three-gluon vertex. In other words: the first case is a three-gluon vertex with one of the gluons interacting with itself, whereas in the latter case, one gluon splits in two, which in turn, exchange another gluon. 

It is not trivial how the cutoff dependence appearing in different diagrams cancels each other exactly in the limit $m_g\to 0$. The running coupling is in fact insensitive to the gluon mass and cutoff, as long as they are sufficiently small. Yet, the presence of the mass is indispensable for the cancellation of divergences.  
The fact that the function $h(x_1)$ appears and that it is different from the functions $h$ obtained in the old version of the calculation~\cite{gomez-rocha_asymptotic_2015} evidences the finite dependence on the regularization procedure, which needs to be clarified and better understood in a deeper analysis.

\section{ Summary and Conclusions }
\label{sec:conclusions}

We have analyzed the effect of a new regularization with canonical gluon mass in the front-form Hamiltonian within the RGPEP. Following previous studies, we have derived the three-gluon-vertex effective Hamiltonian term in a third-order expansion in powers of the coupling constant. 
The inclusion of a gluon mass provides a suitable regularization procedure for small-$x$ divergences. The presence of the mass makes regularization factors go to zero faster than any positive power of $x$ as $x\to 0$.

We have computed the running of the coupling constant using the coefficient function in front of the three-gluon-vertex structure in the approximation $\kappa_{12}^\perp \approx 0$, describing the running of the coupling by its variation with the energy scale. 

Although the new regularization procedure differs conceptually from the one employed in previous calculations, it leads to a similar type of function, exhibiting asymptotic freedom. The running of the coupling depends on the renormalization-group parameter $\lambda$ in the same way as the running coupling calculated using Feynman diagrams for off-shell Green's functions depends on momentum, even though both approaches are fairly different. 
    
In our definition of the running coupling, all terms depending on $m_g$ cancel out in a non-trivial way for infinitesimally small gluon mass. Hence, there is no dependence on the mass parameter $m_g$ or on the cutoff $t_r$, even though separate contributions diverge in this limit. 
  
As in the foregoing studies, a finite dependence on the regularization remains, which is manifest in the appearance of the function $h(x_1)$ in Eq.~(\ref{Rcfinal}). 
For $x_1=0.5$ the running coupling has values extremely close to the one provided by standard calculations (and would be identical for $h(x_1)=0$), with less than a 2\% difference within the entire range of the scale $\lambda$ considered. 
In contrast to the previous calculation, the dependence on $x_1$ appears not only in contribution ($a$), Eq.~(\ref{terma}), but also in $(d)+(f)$, Eq.~(\ref{termdf}). 
Removal of such finite dependence requires a deeper understanding of how to define the running coupling constant in Hamiltonians and, eventually, a higher-order analysis. 
The inclusion of longitudinal polarization for gluon with mass ought to be considered~\cite{glazek-2023}.

\section*{Acknowledgments}
We thank Stanis\l aw D. G\l azek for helpful discussions and acknowledge financial support from the FEDER funds, Project Ref. A-FQM-406-UGR20, from MCIN/ AEI/10.13039/501100011033, Projects Ref. PID2020-114 767GB-I00, PID2019-108655GB-I00, PID2019-106080GB-C21 and FPU21/04180 grant from the Spanish Ministry of Universities.

\appendix

\section{ SOLUTIONS ORDER BY ORDER }
\label{ap:orderbyorder}

This Appendix summarizes the main steps for systematically solving the RGPEP equation~(\ref{eq:rgpep}) perturbatively. The reader interested in a deeper discussion may consult Ref.~\cite{glazek_perturbative_2012}. 
The effective Hamiltonian is a solution to the RGPEP equation: 
\begin{eqnarray}
\cH_t' &=& [[\cH_f,\cH_{Pt}],\cH_t]  \ . 
\end{eqnarray}
The operator $\cH_{Pt}$ is defined in terms of the effective Hamiltonian $\cH_t$, differing in a factor $(\h \sum_i p^+_i)^2$, with $i$ denoting the particles involved in an interaction~\cite{glazek_perturbative_2012}. 

The perturbative solution is given as a power expansion in the coupling constant $g$:
\begin{eqnarray}
\cH_t 
=
\cH_0 + g\cH_{t1} + g^2\cH_{t2} + g^3\cH_{t3} + g^4\cH_{t4} + \dots 
\qquad
\end{eqnarray}
For instance, in a 4th-order expansion one has

\begin{figure}
    \centering
    \includegraphics[width=0.25\textwidth]{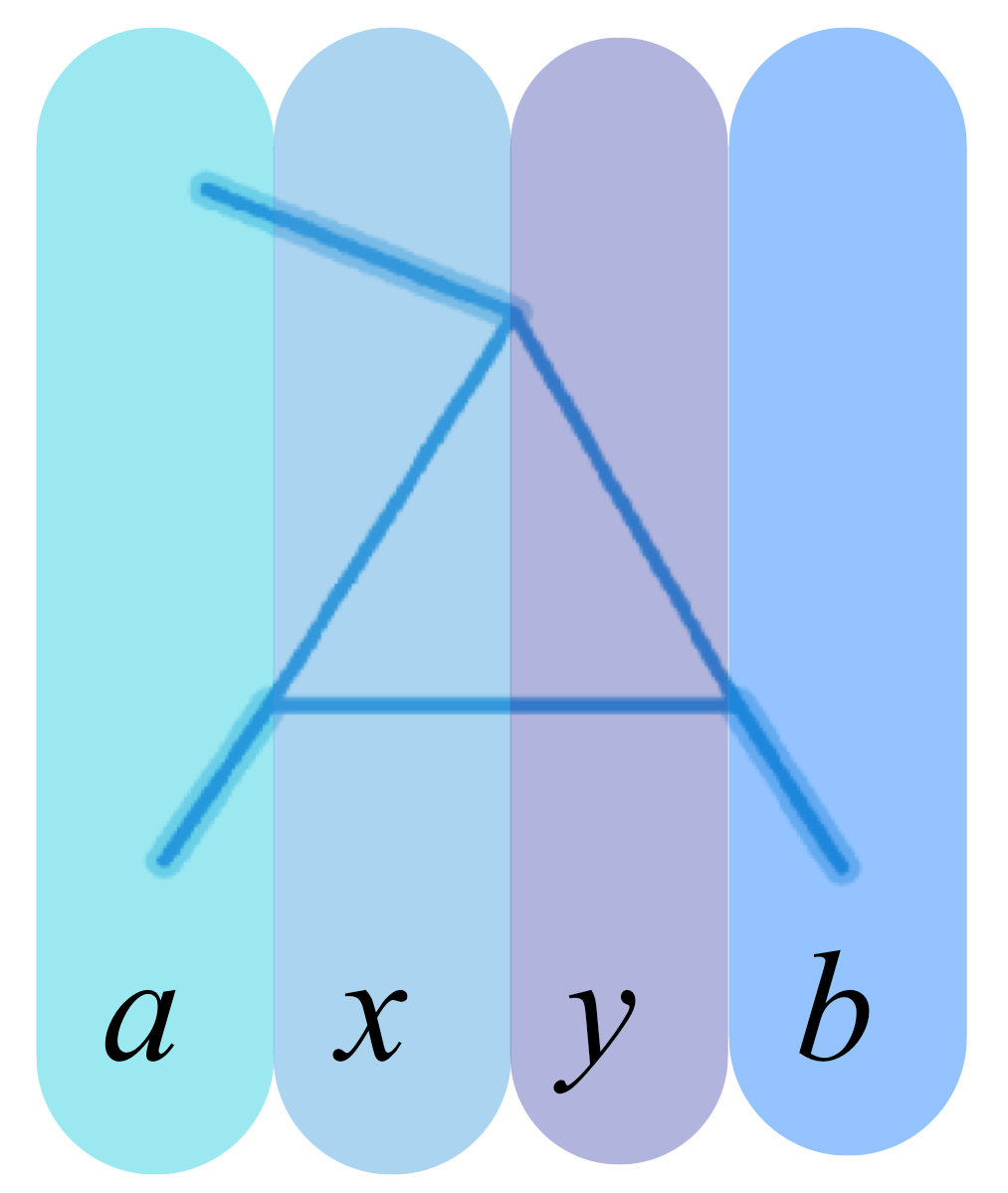}
    \caption{Particle configurations $a$, $x$, $y$, $b$ in diagram $(a)$.}
    \label{fig:configurations}
\end{figure}

\begin{widetext}
    
\begin{eqnarray}
&& \cH_{0}' + g\cH_{t1}' + g^2\cH_{t2}' + g^3\cH_{t3}' + g^4\cH_{t4}' \nn
&=&\left[\left[\cH_0,\cH_0 + g\cH_{1P t} + g^2\cH_{2P t} + g^3\cH_{3P t} + g^4\cH_{4P t} \right],\cH_0 + g\cH_{t1} + g^2\cH_{t2} + g^3\cH_{t3} + g^4\cH_{t4} \right] \ ,
\end{eqnarray}
which can be solved order by order:
\begin{eqnarray}\cH_0' \es 0 \ , \\
g\cH_{t\,1}' 
\es \left[\left[ \cH_0, g\cH_{1P t}\right],\cH_0\right] \ , \\
g^2\cH_{t\,2}' 
\es \left[\left[ \cH_0, g^2\cH_{2P t}\right], \cH_0\right] + \left[\left[ \cH_0, g\cH_{1P t}\right],g \cH_{1 t}\right] \ , \\
g^3 \cH_{t\,3}' 
\es \left[\left[ \cH_0, g^3\cH_{3P t}\right], \cH_0\right] + \left[\left[ \cH_0, g^2\cH_{2P t}\right], g\cH_{1 t}\right] +
\left[\left[ \cH_0, g\cH_{1P t}\right], g^2\cH_{2 t}\right]  \ ,\\
g^4 \cH_{t\,4}' \es \left[\left[ \cH_0, g^4\cH_{4P t}\right], \cH_0\right] + \left[\left[ \cH_0, g^3\cH_{3P t}\right], g\cH_{1 t}\right] +
\left[\left[ \cH_0, g^2\cH_{2P t}\right], g^2\cH_{2 t}\right] + \left[\left[ \cH_0, g\cH_{1P t}\right], g^3\cH_{3 t}\right] \ .
\end{eqnarray}
We present solutions to these equations in terms of matrix elements $\cH_{t\,ab}=\langle a|\cH|b\rangle$.
\begin{eqnarray}
\cH_{t1\,ab} 
\es
f_{t\,ab}\,\cH_{01\,ab}  \ ,
\\
\cH_{t2\,ab} 
\es  f_{t\,ab}\, \sum_x
\cH_{01\,ax}\cH_{01\,xb}\, \cB_{t\,axb}
\ + \ f_{t\,ab}\,\cG_{02\,ab}  \ ,
\\
\cH_{t3\,ab} 
\es 
f_{t\,ab}\,\cG_{03\,ab}  +  
f_{t\,ab}\,
\sum_{xy}\cH_{01\,ax}\,\cH_{01\,xy}\,\cH_{01\,yb} \, \cC_{t\,axyb}
\, + \,
f_{t\,ab}\,\sum_{x} 
\,
\left(
\cH_{01\,ax}\,\cG_{02\,xb}
 +  \cG_{02\,ax}\,\cH_{01\,xb}
\right)\,  \cB_{t\,axb}  \ , \\
\cH_{t4\,ab} 
\es 
f_{t\,ab}\,\sum_{xyz} \cH_{01\,ax}\,
\cH_{01\,xy}\,\cH_{01\,yz}\,\cH_{01\,zb} \,\,\cD_{t\,axyzb}
\np
f_{\tau\,ab}\,\sum_{xy} \left(\cH_{01\,ax}\,
\cH_{01\,xy}\,\cG_{02\,yb}
\ + \ \cH_{01\,ax}\,\cG_{02\,xy}\,\cH_{01\,yb}\ + \ \cG_{02\,ax} \,\cH_{01\,xy}\cH_{01\,yb}
\right)
\, \cC_{t\, axyb}
\np 
f_{\tau\,ab}\,\sum_x  \left(\cG_{02\,ax} 
\,\cG_{02\,xb}+ \cH_{01\,ax}\,\cG_{03\,xb} + \cG_{03\,ax}\, \cH_{01\,xb}\right)
\, \cB_{t\, axb}
\ + \ f_{\tau\,ab}\, \cG_{04\,ab} \ ,
\end{eqnarray}
were operators $\cG_{02}$, $\cG_{03}$, $\cG_{04}$ are the initial condition, being canonical terms or counterterms of the respective order~\cite{glazek_perturbative_2012}. The so-called RGPEP factors are given by:
\begin{eqnarray}
\cA_{t\,axb} 
\es
\left[ax\, p_{ax}\, + bx\, p_{bx}\right] f_{t\,ab}^{-1}f_{t\,ax}f_{t\,bx} \ , 
\\
\cB_{t\,axb} 
\es  
\int_0^t \cA_{\tau\,axb}\, d\tau \ , 
\\
\cC_{t\,axyb} 
\es 
\int_0^t\left[\cA_{\tau\,axb} 
  \,\cB_{\tau\,xyb} \ + \ 
\cA_{\tau\,ayb}\,\cB_{\tau\,axy}\right] d\tau \ , \\
\cD_{t\,axyzb} \ ,
\es
\int_0^t \left[\cA_{\tau\,axb}\, \cC_{\tau\,xyzb} 
\ + \
\cA_{\tau\,azb}\,
 \cC_{\tau\,axyz} 
\ + \ 
A_{\tau\,ayb}\,
\, \cB_{\tau\,axy} \,
\,\cB_{\tau\,yzb} \right] \ ,
\end{eqnarray}
with $f_{t\, ab}:=\exp\left[-ab^2\, t\right]$ and 
$ab:=\cM_{ab}^2-\cM_{ba}^2$ .

The subscripts, $a$, $b$, $x$, and $y$ represent particle configurations, cf. Figure~\ref{fig:configurations} and Ref.~\cite{glazek_perturbative_2012}.
In this paper we consider solutions up to third order. 

\end{widetext}

\section{INTEGRATION PROCEDURE}
\label{integration}

In this appendix a systematic procedure to integrate expressions such as Eq.~(\ref{gamma_a_integral}) is discussed. After introduction of the scale $t_{0}$ integrals involved have the structure
\begin{eqnarray}
&&\lim_{m_{g}\rightarrow0} \int_{1-x_{P}}^{1} dx \int_{0}^{\infty} d^2 \kappa^{\perp} 
\nt 
\tilde f_{t_{r}}\left(f_{t,i}-f_{t_{0},i}\right)C_{i}\left(x,x_{1},\kappa,\kappa_{12},m_{g}\right),
\label{general_integral}
\end{eqnarray}
where $\tilde f_{t_{r}}$ is the product of regularization form factors introduced at each interaction vertex and $\left(f_{t,i}-f_{t_{0},i}\right)C_{i}$ denotes a sum of differences of form factors, evaluated at $t$ and $t_{0}$, multiplied by fractions depending on invariant masses [see $\mathcal{B}_{t(a)}$ in Eq.~(C15) of~\cite{gomez-rocha_asymptotic_2015}, where the counterterm has not been taken into account yet]. $x_{P}$ denoted the minimal momentum fraction carried in internal loops. In this work, it is equal to 1 or $x_{2}$.

The limit $m_{g}\rightarrow0$ is used to simplify the integrand. First, we introduce dimensionless variables using an arbitrary scale $t_{N}$:
\begin{equation}
\bar{t}=\frac{t}{t_{N}} \ , \ \bar{\kappa}^{\perp}=\kappa^{\perp}t_{N}^{1/4}   \ , \  \bar{m}_{g}=m_{g} t_{N}^{1/4} \ .
\label{dimensionless}
\end{equation}
After this change of variables, the behavior of the integrand in the neighborhood of $\kappa = 0$ and $x=1$ or $x=1-x_{P}$ is studied. Form factors are decreasing exponentials of differences of invariant masses:
\begin{equation}
    f_{t} = \exp{\left[-t(\cM^2-m_g^2)^2\right]},
\end{equation}
with $\cM$ depending on $\bar{\kappa}, \bar{m}_{g}$ and  $x$ in the following way: 
\begin{equation}
\cM^2 \ = \ 
x_P^2\frac{\bar{\kappa}^2+\bar{m}_{g}^2}{(1-x)\left(x-(1-x_{P})\right)} \ .
\end{equation}
There are two scales settled by $\bar{m}_{g}$ and $\bar{\kappa}$; one corresponds to $\bar{\kappa} >> \bar{m}_{g}$, in which the mass parameter can be ignored, and the other to $\bar{\kappa}\approx0$, for which $\bar{m}_{g}$ provides the necessary regularization when $x$ is close to of $1-x_{P}$ or 1. Thus the limit $\bar{m}_{g}\rightarrow 0$ cannot be applied inside the integral, as it cuts off the divergences of invariant masses that take place when $x-(1-x_{P}) < \bar{m}_{g}$ or $1-x < \bar{m}_{g}$ if $\bar{\kappa}=0$, necessary to regulate \ref{general_integral}.  

If the integral over $x$ is divided in three regions:
\begin{itemize}
    \item 
    Region $\mathbf{I}$: $x\in\left(1-x_{P},(1-x_{P})+\bar{m}_{g}\right)$ \ ,
    \item 
    Region $\mathbf{II}$:  $x\in\left((1-x_{P})+\bar{m}_{g},1-\bar{m}_{g}\right)$ \ ,
    \item 
    Region $\mathbf{III}$: $x\in\left(1-\bar{m}_{g},1\right)$ \ ,
\end{itemize}
it can be greatly simplified in intervals $\mathbf{I}$ and $\mathbf{III}$ applying approximations $1-x\approx x_{P}$ and $x-\left(1-x_{P}\right)\approx x_{P}$ respectively. The limit $\bar{m}_{g}\rightarrow0$ can be safely applied in region $\mathbf{II}$ , since numerators of invariant masses are quadratic in $\bar{m}_{g}$, whereas denominators are linear in $\bar{m}_{g}$.

For example, in the self-energy integral Eq.~(\ref{mut}) the three regions are
\begin{eqnarray}
    \int_0^1 dx  \ \to \  \lim_{\bar m_g \to 0} \left[ \int_0^{\bar m_g} dx + \int_{\bar m_g}^{1- \bar m_g} dx + \int_{1- \bar m_g}^1 dx \right] \ .
\end{eqnarray}
The contribution in region $\mathbf{I}$ is obtained using the approximation $1-x\approx 1$:
\begin{equation}
\mu_{t}|_{\mathbf{I}}=\frac{N_{c}}{\left(2\pi\right)^{2}}
\int_{0}^{\bar{m}_{g}}
\frac{dx}{x^{2}}
\int d\bar{\kappa}\frac{\bar{\kappa}^{3}}{\bar{m}_{g}^2+\bar{\kappa}^2}
\exp\left[-2\bar{t}\frac{(\bar{m}_{g}^2+\bar{\kappa}^2)^{2}}{x^{2}}\right],
\end{equation}
that can be easily integrated:
\begin{eqnarray}
\mu_{t}|_{\mathbf{I}}
\es -\frac{N_{c}}{\left(2\pi\right)^{2}}\sqrt{\frac{\pi}{2}}\frac{1}{8\sqrt{t}}\left[2+\gamma_{E}+3\log\left(2\right)+\log\left(tm_{g}^{4}\right)
\right.
\nm 
\left.
\frac{1}{2}\log\left(t_{N}m_{g}^{4}\right)\right].
\end{eqnarray}
In region $\mathbf{III}$ the change of variables $y = 1-x$ makes the integral analogous to the one on interval region $\mathbf{I}$ and the result is the same. In region $\mathbf{II}$ the regularization can be safely lifted by taking the limit $t_r \to 0$ and $m_g\to 0$, yielding
\begin{eqnarray}
\mu_{t}|_{\mathbf{II}}
\es 
\frac{N_{c}}{\left(2\pi\right)^{2}}\int_{\bar{m}_{g}}^{1-\bar{m}_{g}}dx\int d\kappa\kappa\left[1+\frac{1}{x^{2}}+\frac{1}{\left(1-x\right)^{2}}\right]
\nt 
\exp\left[-2t\frac{\kappa^{4}}{x^{2}\left(1-x\right)^{2}}\right] 
\nn
\es  
-\frac{N_{c}}{\left(2\pi\right)^{2}}\sqrt{\frac{\pi}{2}}\frac{1}{8\sqrt{t}}\left[\log\left(t_{N}m_{g}^{4}\right)+\frac{11}{3}\right] \ .
\end{eqnarray}
Adding the results of the three intervals one obtains  Eq.~(\ref{self-energy result}). The dependence on the arbitrary scale $t_{N}$ cancels and a logarithmic dependence on $m_{g}$ is left as a remnant of the regulated divergence.

The error induced by this approximation is negligible in the limit of infinitesimally small gluon mass. This is shown in Figure~\ref{fig:my_label}: The left panel shows the difference between the result obtained applying this procedure and the numerical integral, as a function of the gluon mass parameter $m_g$. The right panel shows the difference between the approximated analytical integral and the numerical one for different values of $t$ and a fixed mass $m_g=0.01$ MeV.
We recall that we are interested in the limit $m_g \to 0$.

\begin{widetext}
    
\begin{figure}
    \centering
    \includegraphics[scale = 0.6]{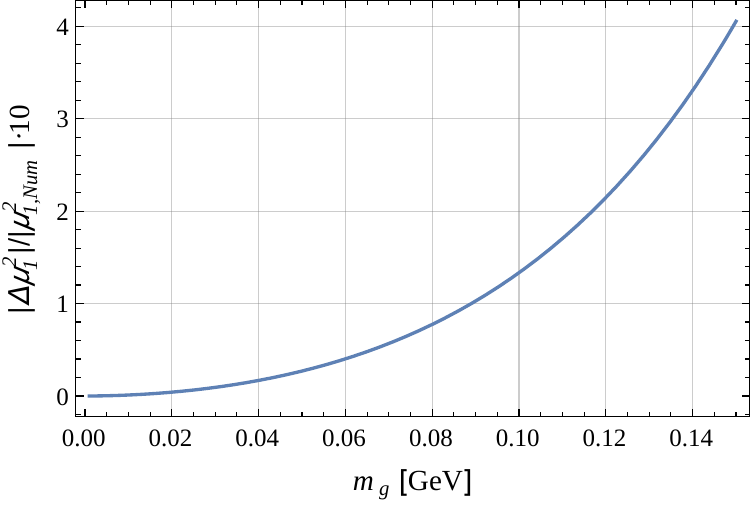}
    \quad \includegraphics[scale = 0.6]{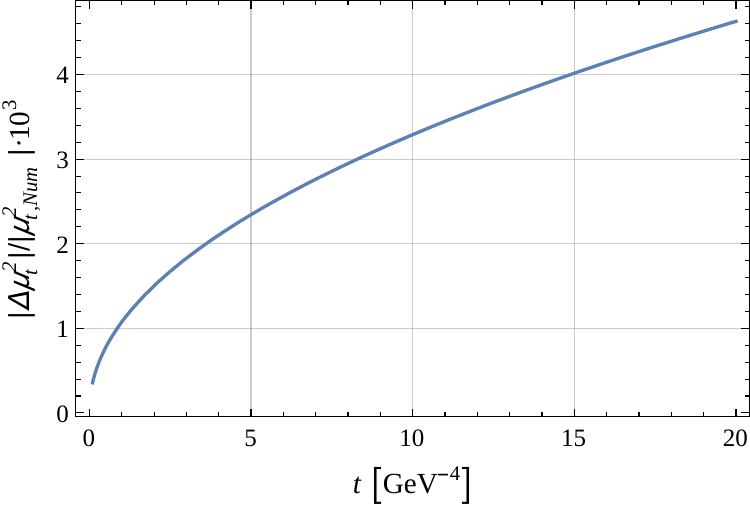}
    \caption{Numerical check of the approximation taken in the integral of Eq.~(\ref{mut}). The difference $\Delta\mu^{2}_{t}=\mu^{2}_{t,\text{Num}}-\mu^{2}_{t,\text{App}}$ normalized to the value of the numerical integration $\mu^{2}_{t,\text{Num}}$ is plotted as function of $m_{g}$ for fixed $t$ and vice-versa: In the left figure, $t = 1 \text{ GeV}^{-4}$ and $m_{g}$ goes from $10^{-3}$ to 0.15 GeV; in the right plot $m_{g} = 0.01$ GeV and $t$ goes from 0.1 to 20 $\text{GeV}^{-4}$. The result is exact in the limit $m_g\to 0$.}
    \label{fig:my_label}
\end{figure}

\end{widetext}

\bibliography{AF2023}

\bibliographystyle{apsrev4-2}

\end{document}